\documentclass[aps,pre,twocolumn,showpacs]{revtex4}
\usepackage[dvips]{graphicx}
\usepackage{amsmath}
\usepackage{times}

\begin{document}
\author{%
Dmitry V. Savin$^{1,2}$ and Hans-J\"urgen Sommers$^1$}
\affiliation{%
$^1$Fachbereich Physik, Universit\"at Duisburg-Essen, 45117 Essen,
Germany}
\affiliation{%
$^2$Budker Institute of Nuclear Physics, 630090 Novosibirsk, Russia}

\title{Delay times and reflection in chaotic cavities with absorption}
\date{July 12, 2003 }%
\published{: Phys. Rev. E \textbf{68} (2003) to appear}

\begin{abstract}
Absorption yields an additional exponential decay in open quantum systems
which can be described by shifting the (scattering) energy $E$ along the
imaginary axis, $E+i\hbar/2\tau_{a}$. Using the random matrix approach, we
calculate analytically the distribution of proper delay times (eigenvalues of
the time-delay matrix) in chaotic systems with broken time-reversal symmetry
that is valid for an arbitrary number of generally nonequivalent channels and
an arbitrary absorption rate $\tau_{a}^{-1}$. The relation between the
average delay time and the ``norm-leakage'' decay function is found.
Fluctuations above the average at large values of delay times are strongly
suppressed by absorption. The relation of the time-delay matrix to the
reflection matrix $S^{\dagger}S$ is established at arbitrary absorption that
gives us the distribution of reflection eigenvalues. The particular case of
single-channel scattering is explicitly considered in detail.
\end{abstract}



\maketitle

There is a growing interest in statistical properties of the Wigner-Smith
matrix $Q(E)$=$-i\hbar S^{\dagger}\partial S/\partial E$
\cite{Wigner1955,Smith1960}, with $S(E)$ being the scattering matrix at the
collision energy $E$, in the cases of chaotic scattering and transport in
disordered media \cite{review}. In the resonance scattering, the matrix
element $Q_{cc'}$ describes the overlap of the internal parts of the
scattering wave functions in the incident channels $c$ and $c'$
\cite{Smith1960,Sokolov1997}. This directly relates the Wigner-Smith matrix
to the effective non-Hermitian Hamiltonian
$\mathcal{H}=H-\frac{i}{2}VV^{\dagger}$ of the unstable intermediate system
as follows (henceforth $\hbar$=1) \cite{Sokolov1997}:
\begin{equation}\label{Q}
Q(E) = V^{\dagger}\frac{1}{(E-{\cal H})^{\dagger}}\,
     \frac{1}{E-{\cal H}}V \,.
\end{equation}
The Hermitian part $H$  stands here for the closed counterpart of the system
while the amplitudes $V_n^c$ describe the coupling between $N$ interior and
$M$ channel states. The random-matrix theory approach is usually adopted
to simulate the complicated intrinsic motion
\cite{Verbaarschot1985,Beenakker1997,Alhassid2000}.

The known analytical results
\cite{Lyuboshitz1977,Lehmann1995b,Fyodorov1997,Gopar1996,Fyodorov1997i,%
Brouwer1997,Sommers2001} are restricted to the idealization neglecting
absorption. The latter is, however, always present to some extent under
laboratory conditions, being one of the sources of a coherence loss in
quantum transport. This has dramatic consequences for the statistical
observables \cite{Doron1990,Huibers1998b}. Necessity of proper accounting of
finite decoherence \cite{Buettiker1986} was recently emphasized
\cite{Alves2002} in order to remove a certain discrepancy between theory
\cite{Baranger1995,Brouwer1997ii} and experiment \cite{Huibers1998b} on
conductance distributions in quantum dots. Reflection in a weakly absorbing
medium turned out to be directly related
\cite{Doron1990,Ramakrishna2000,Beenakker2001} to the time-delay matrix
without absorption. Recent experiments \cite{Schaefer2003} in microwave
cavities demonstrated that the absorption (due to the skin effect in the
walls) may be strong, leading to an exponential decay
\cite{Doron1990,Schaefer2003}.

In this paper we show that representation (\ref{Q}) in terms of the effective
Hamiltonian allows us to extent the consideration to the case of an arbitrary
absorption. The nature of the exponential decay caused by absorption can be
easily understood from the following model consideration which actually goes
back to the concept of the spreading width in nuclear physics
\cite{Bohr1969}; see \cite{Sokolov1997} for the recent developments. In
addition to coupling to continuum (scattering) states the originally closed
system is considered to be also coupled to the background compound
environment. The latter has a very dense spectrum with the mean level spacing
$\Delta_{\rm bg}$ being much smaller than the corresponding one $\Delta$ of
the closed system, $\Delta_{\rm bg}\ll\Delta$. When the coupling strength
$v^2$ is large enough to mix background states, $v^2>\Delta^2_{\rm bg}$, the
original levels acquire the damping or spreading width
$\Gamma_{\downarrow}\equiv2\pi v^{2}/\Delta_{\rm{bg}}$
\cite{Bohr1969,Sokolov1997}. Corrections to the resulting exponential decay
show up at the time $t_{*}\sim\Delta^{-1}_{\rm bg}$ and, therefore, can be
safely ignored on mesoscopic scale of the Heisenberg time
$t_H\equiv2\pi/\Delta\ll t_{*}$ we are interested in.

\begin{figure}[hb]
\includegraphics[width=0.465\textwidth]{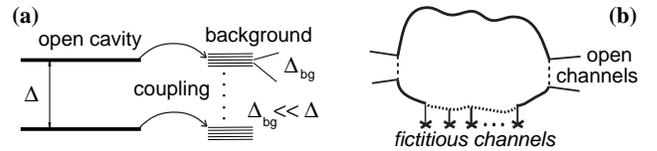}
\caption{An open cavity with absorption in walls modelled by
coupling to (a) the compound background or (b) an infinite number
of fictitious channels with vanishing transmission probabilities.}
\end{figure}
In the absorption limit of continuous spectrum of the background, when an
irreversible decay into walls takes place, this description becomes
equivalent to that achieved in the framework of the B\"uttiker's model of
dephasing in mesoscopic conductors; see Fig.~1 \cite{models}. One considers
\cite{Buettiker1986,Baranger1995} $M_{\phi}$ fictitious scattering channels
in addition to $M$ real ones. The vanishing transmission $T_{\phi}\to0$ of
the fictitious channels is assumed to be compensated by their large number
$M_{\phi}\to\infty$, the dimensionless absorption rate
$\gamma=M_{\phi}T_{\phi}$ being kept fixed \cite{Brouwer1997ii}. Then the
anti-Hermitian part of the effective Hamiltonian $\mathcal{H}$, which
describes coupling to (all) open channels, splits readily as
$\sum^{M}_{c,{\rm\,real}}V^c_nV^{c\,*}_{m}+\delta_{nm}\Gamma_{a}$
\cite{Brouwer1997ii} into the escape contribution (first term) and damping
one with $\Gamma_{a}\equiv\tau_{a}^{-1}\equiv\gamma\Delta/2\pi$. An
associated with the last term exponential decay lasts up to the
characteristic time $t_{*}=t_{H}/\sqrt{\gamma T_{\phi}}$ \cite{Savin1997}
being large as compared to $t_{H}$.

The consideration presented suggests  that \emph{nonzero} absorption is
equivalent to the purely imaginary shift $E+\frac{i}{2}\Gamma_{a}\equiv
E_{\gamma}$ of the energy in the Green's function $(E-\mathcal{H})^{-1}$ of
the open system \emph{without} absorption as long as resonance scattering far
from the channel thresholds is concerned \cite{shift}; see also
\cite{Beenakker2001,Kogan2000}. This is in agreement with available data on
correlations of $S$ matrix elements in cavities with absorption
\cite{Schaefer2003}.

In what follows we consider the time-delay matrix with absorption
$Q_{\gamma}\equiv Q(E_{\gamma})$, with $Q$ from (\ref{Q}), treating
$\gamma\!=\!\Gamma_{a}t_H$ as a phenomenological parameter. The important
relation for the reflection matrix
\begin{equation}\label{R}
R \equiv S_{\gamma}^{\dagger}S_{\gamma} = 1-\Gamma_{a}Q_{\gamma}
\end{equation}
follows directly from the definition of the scattering matrix
$S_{\gamma}\equiv
S(E_{\gamma})=1-iV^{\dagger}(E_{\gamma}-\mathcal{H})^{-1}V$, which is
subunitary ($R<1$) at nonzero absorption. This relation gives $Q_{\gamma}$
the meaning of the matrix of unitarity deficit and generalizes limiting
expressions of Refs.~\cite{Ramakrishna2000,Beenakker2001} valid at weak
absorption to the case of \textit{arbitrary} $\Gamma_{a}$. $Q_{\gamma}$ is a
$M\!\times\!M$ Hermitian, positive-definite matrix and, therefore, has real
positive eigenvalues $q_c$, the so-called \emph{proper} delay times. They
were recently studied in much detail for the case of zero absorption
\cite{Brouwer1997,Sommers2001}. Even a weak absorption modifies their
statistical properties significantly, as will be shown below.

We begin with the calculation of the average total delay time
$q_\mathrm{tot}\equiv\overline{q_1+\cdots+q_M}=\mathrm{tr\,}\overline{Q_{\gamma}}$,
where the bar denotes the ensemble average. Making use of the invariance of
the trace under cyclic permutations and the following relation
$VV^{\dagger}=i[(E_{\gamma}-\mathcal{H})^{\dagger}-
(E_{\gamma}-\mathcal{H})]-\Gamma_{a}$, one gets
\begin{equation}\label{q_w}
q_\mathrm{tot} =
\mathrm{Im\,}\mathrm{Tr\,}\overline{\frac{-2}{E_{\gamma}-\mathcal{H}} } -
\Gamma_{a}\mathrm{Tr}\overline{\left( \frac{1}{ E_{\gamma}-\mathcal{H} }
\frac{1}{(E_{\gamma}-\mathcal{H})^{\dagger} }\right) }\,,
\end{equation}
where $\mathrm{Tr}$ acts in the $N$-dimensional intrinsic space of
resonances. The first term is known \cite{Lyuboshitz1977,Lehmann1995b} to be
equal to the Heisenberg time $t_H$. To calculate the second one, it is
instructive to go to the time domain and to exploit the well-known relation
between the Green's function and the time evolution operator
$\exp(-i\mathcal{H}t)$. This enables us to represent (\ref{q_w}) in the
following form:
\begin{equation}\label{t-P}
\tau_\mathrm{tot} \equiv \frac{ q_\mathrm{tot} }{ t_H } = 1 - \Gamma_{a}
\int_0^{\infty}dt\,e^{-\Gamma_{a}t}P(t)\,,
\end{equation}
where
$P(t)\equiv(1/N)\mathrm{Tr}\overline{(e^{i\mathcal{H}^{\dagger}t}e^{-i\mathcal{H}t})}$
is the ``norm-leakage'' decay function introduced in Ref.~\cite{Savin1997}.
The average delay time within the cavity becomes smaller due to additional
dissolution in the walls. The average weighted-mean reflection coefficient
$\langle{r}\rangle\equiv M^{-1}\mathrm{tr}\overline{R}$ is correspondingly
given by  $\langle{r}\rangle=1-\gamma\tau_\mathrm{tot}/M$.

$P(t)$ can be calculated by means of Efetov's supersymmetry technique
\cite{Efetov1996,Verbaarschot1985}, which becomes now a standard analytical
tool. Here we only state the corresponding result for the case of preserved
time-reversal symmetry (TRS):
\begin{eqnarray}\label{P}
P(t)&=& \int_{-1}^{1}\!\!d\lambda \int_{1}^{\infty}\!\!d\lambda_1
\int_{1}^{\infty}\!\!d\lambda_2 \, \mu(\lambda_i) \,
\delta\Bigl(\frac{t}{t_H}-\frac{\lambda_1\lambda_2\!-\!\lambda}{2}\Bigr)
\nonumber \\
&&\times f(\lambda_i)\prod_{c=1}^{M}\left[
\frac{(g_c+\lambda)^2}{(g_c\!+\!\lambda_1\lambda_2)^2-
(\lambda_1^2\!-\!1)(\lambda_2^2\!-\!1)}\right]^{\frac{1}{2}}\!,\
\quad
\end{eqnarray}
where
$\mu(\lambda_i)=(1-\lambda^2)/(\lambda_1^2+\lambda_2^2+\lambda^2-2\lambda\lambda_1\lambda_2-1)^2$
and
$f(\lambda_i)=(2\lambda_1^2\lambda_2^2-\lambda_1^2-\lambda_2^2-\lambda^2+1)/4$.
The quantities $g_c=2/T_c-1\geq1$ are related to the transmission
coefficients $T_c=1-|\overline{S_{cc}}|^2$ \cite{Verbaarschot1985}, which
determine the openness strength of the system (without absorption), referring
$T=1$ (0) to the completely open (closed) one. For reader's convenience, we
note that the result for the case of broken TRS \cite{Savin1997} follows from
(\ref{P}) by removing there the $\lambda_2$ integration and setting
$\lambda_2\!=\!1$ everywhere in the integrand save the integration measure
$\mu(\lambda_i)\!=\!(\lambda_1-\lambda)^{-2}$ in this case \cite{misprint}.
It is worth also pointing out the relation between $P(t)$, Eq.~(\ref{P}), and
the autocorrelation function of the photodissociation cross section
\cite{Fyodorov1998i}. The exact (in the RMT limit $N\to\infty$) equation
(\ref{t-P}) is valid for any symmetry and will be also derived below using a
different way.

The ``norm-leakage'' is identical to unity when the system is closed (hence
the norm). Its time dependence is solely due to the openness of the system
and has been thoroughly studied in \cite{Savin1997} that allows us to
understand the qualitative dependence of $q_\mathrm{tot}$ on absorption. The
typical behavior $P(t)\sim\prod_{c=1}^{M}[1+(2/\beta)T_c t/t_H]^{-\beta/2}$,
with $\beta$$=$1(2) standing for preserved (broken) TRS, is the simple
exponential $\exp(-t\sum_cT_c/t_H)$ at small enough times. In the so-called
``diagonal approximation'' \cite{Savin1997}, which neglects the
nonorthogonality of the resonance wave functions and becomes asymptotically
exact at large $t$, $P(t)$ turns out to be related by the Laplace transform
$P_{\rm diag}(t)= \int_{0}^{\infty}\!\!d\Gamma\,e^{-\Gamma t}
\rho(\Gamma)\equiv\langle e^{-\Gamma t}\rangle_{\Gamma}$ to the distribution
$\rho(\Gamma)$ of resonance widths. One gets readily from (\ref{t-P}) that
$\tau_\mathrm{tot}=
\left\langle\Gamma/(\Gamma+\Gamma_{a})\right\rangle_{\Gamma}$ within this
very approximation. The simple interpolation formula
$\tau_\mathrm{tot}\approx (1+\gamma/\sum_cT_c)^{-1}$ with corrections of the
order of $\mathrm{min}[1/\gamma,1/\sum_cT_c]$ becomes exact as the absorption
rate $\gamma$ and/or the total (dimensionless) escape width $\sum_cT_c$
grows.

We proceed  further with an analysis of the distribution of the proper delay
times $\mathcal{P}(q)=M^{-1}\sum_c\overline{\delta(q-q_c)}$. For the sake of
simplicity, we restrict ourselves to the case of broken TRS (the unitary
symmetry class). The factorized representation (\ref{Q}) of $Q_{\gamma}$
enables us to use the same  method developed in Ref.~\cite{Sommers2001} to
treat the zero absorption case. Thus, we skip all standard technical details,
indicating only essential ones. As usual, the jump of the resolvent
$G(z)=M^{-1}\mathrm{tr\,}\overline{(z-Q_{\gamma})^{-1}}$ on the discontinuity
line along $q=\mathrm{Re}z>0$ determines the seeking distribution as follows:
$\mathcal{P}(q)=\pi^{-1}\mathrm{Im\,}G(q\!-\!i0)$. Due to the factorized
structure of $Q_{\gamma}$, $G(z)$ can be then represented in the form
suitable for subsequent supersymmetry calculation \cite{susy}. We find the
following expression for the determining part $K(\zeta\!=\!z/t_H)\equiv
M\zeta^2(t_HG(z)-1/\zeta)$:
\begin{eqnarray}\label{tildeG}
K(\zeta) &=& 1 +\frac{1}{2} \int^{\infty}_{1}\!\! d\lambda_1
\!\int^{1}_{-1}\!\frac{d\lambda}{\lambda_1-\lambda}
\prod^{M}_{c=1}\frac{g_c+\lambda}{g_c+\lambda_1} \nonumber \\
&&\times \left(\frac{\partial}{\partial\nu_1} -
\frac{\partial}{\partial\nu}\right)
b_{\gamma}(\lambda_1)\,f_{\gamma}(\lambda)\Bigm|_{\nu_1=\nu=1}.
\end{eqnarray}
Here $b_{\gamma}(\lambda_1)$$=$$e^{(1-\gamma\zeta/2)\nu_1\lambda_1/\zeta}
I_0[\frac{\nu_1}{\zeta}\sqrt{(1-\gamma\zeta)(\lambda_1^2-1)}]$ and
$f_{\gamma}(\lambda)$$=$$e^{-(1-\gamma\zeta/2)\nu\lambda/\zeta}
J_0[\frac{\nu}{\zeta}\sqrt{(1-\gamma\zeta)(1-\lambda^2)}]$, with $I_0(x)$
[$J_0(x)$] being the modified (usual) Bessel function. The resolvent
$G(\zeta)$ given by  Eq.~(\ref{tildeG}) is an analytical function of the
complex variable $\zeta$ for the negative values of $\mathrm{Re\,}\zeta$ and,
therefore, can be expanded there in Taylor's series. One finds directly from
the definition of $G$ that
$t_HG(z)=1/\zeta+\overline{\mathrm{tr\,}{Q_{\gamma}}}/(M\zeta^2t_H)+\cdots$
for large $\zeta$, relating thus $q_w$ to the coefficient of the second term
of this expansion. On the other hand, this coefficient is given just by
$K(-\infty)$ which can easily be calculated from (\ref{tildeG}) to reproduce
exactly equation (\ref{t-P}).

An analytical continuation in (\ref{tildeG}) to the region of positive
$\tau\equiv\mathrm{Re\,}\zeta$ requires more care as compared to the case
\cite{Sommers2001} of zero absorption, where it was achieved by a proper
deformation of an original integration contour. First we make the following
decomposition in partial fractions:
$$
\frac{1}{\lambda_1-\lambda} \prod^{M}_{c=1}\frac{g_c+\lambda}{g_c+\lambda_1}
=\frac{1}{\lambda_1-\lambda} - \sum_{a=1}^{M}\frac{1}{g_a+\lambda_1}
\prod_{b(\neq a)}\frac{g_b+\lambda}{g_b-g_a}.
$$
The contribution from the term $(\lambda_1\!-\!\lambda)^{-1}$ leads to an
exact cancellation with the first term in (\ref{tildeG}). [This is not
surprising since the product term in (\ref{tildeG}), the channel factor,
which determines solely the strength of system openness, reduces at
$\lambda_1$=$\lambda$  to unity, resulting $G(z)\!=\!1/z$ in this case
identically.] The integration over $\lambda_1$ gets completely decoupled from
that over $\lambda$ in the contribution from the rest sum. Making use of the
table integrals \cite{table}, one finds that
\begin{equation}\label{integral}
\int_{1}^{\infty}\!\!\frac{d\lambda_1e^{s\lambda_1}}{g+\lambda_1}
I_0(\alpha\sqrt{\lambda_1^2-1}) = \int_{0}^{\infty}\!\!dp
\frac{e^{-gp}e^{\sqrt{(p-s)^2-\alpha^2}}}{\sqrt{(p-s)^2-\alpha^2} },
\end{equation}
with shorthands $s\equiv\tau^{-1}-\gamma/2$ and
$\alpha\equiv\tau^{-1}\sqrt{1-\tau\gamma}$. Just this term (\ref{integral})
has a nonzero imaginary part, thus the distribution, at positive $\tau-i0$. A
close inspection of the r.h.s. of Eq.~(\ref{integral}) shows, that the
imaginary part is determined by the integration region $s-\alpha<p<s+\alpha$,
resulting at the end in
$\pi{I}_0(\alpha\sqrt{g^2-1})\Theta(\tau^{-1}-\gamma)$, with the step
function $\Theta(x)$.

We arrive finally at the following general expression for the probability
distribution of the proper delay times:
\begin{equation}\label{Ptau}
\mathcal{P}(\tau=\frac{q}{t_H}) = \frac{1}{M} \sum^M_{c=1}
\left(\frac{\partial}{\partial\nu}-\frac{\partial}{\partial\nu_1}\right)
B_cF_c\Bigm|_{\nu_1=\nu=1} \,,
\end{equation}
for $0<\tau\leq \gamma^{-1}$, and $\mathcal{P}(\tau)\!\equiv\!0$ otherwise.
Here
\begin{subequations}\label{BF}
\begin{eqnarray}
&& B_c = e^{-\nu_1sg_c}\,I_0(\nu_1\alpha\sqrt{g_c^2-1})
    \prod_{a (\neq c)}{1\over g_a -g_c}                  \label{B}\,, \\
&& F_c = \int_{-1}^{+1}\!\frac{d\lambda}{2} e^{-\nu s\lambda}\,
   J_0(\nu\alpha\sqrt{1-\lambda^2})
   \prod_{a (\neq c)}{(g_a+\lambda)}. \quad\quad   \label{F}
\end{eqnarray}
\end{subequations}
The obtained result is valid for arbitrary absorption strength and
arbitrary transmission coefficients of $M$ generally nonequivalent channels.
The limit of zero absorption \cite{Sommers2001} is correctly reproduced. The
case of statistically equivalent channels can easily be worked out by
performing the limiting transition $g_c\to g=2/T-1$ for all $c$. At last,
the distribution function $P_R(r)=M^{-1}\sum_{c}\overline{\delta(r-r_c)}$ of
\emph{reflection eigenvalues} $r_c=1-\gamma q_c/t_H$
follows readily from (\ref{Ptau}) as
\begin{equation}\label{P_R}
P_R(r) = \gamma^{-1}\mathcal{P}\bigl[\gamma^{-1}(1-r)\bigr]\,,\quad 0\leq r<1 .
\end{equation}

We see that the absorption rate $\gamma$ enters the distribution in a highly
nontrivial way. This is expected to be true for any distribution function and
is contrasted with a correlation function of, say, $S$ matrix elements
$\overline{S^{\ast}_{ab}(E_{\gamma})S_{a'b'}(E_{\gamma}+\varepsilon)}$. The
corresponding form-factor \cite{Schaefer2003} (the Fourier transform of the
correlation function) differs from that \cite{Verbaarschot1985} of the zero
absorption case simply by the presence of an additional exponential term
$e^{-\gamma t/t_H}$. The most striking effect of finite absorption on the
time-delay distribution is likely to consist in suppression of the universal
long-time tails $\tau^{-M\beta/2-2}$
\cite{Fyodorov1997,Gopar1996,Fyodorov1997i,Brouwer1997,Sommers2001}  at
$\tau>\gamma^{-1}$ \cite{tail}. To understand this fact qualitatively, we
note that the delay time
$q(E)\approx\Gamma_n/[(E-E_n)^2+\frac{1}{4}(\Gamma_n+\Gamma_{a})^2]$ in a
vicinity of a given resonance with the energy $E_n$. The maximal value of
this single-resonance contribution is attained at $E=E_n$, being $q_{\rm
max}=4\Gamma_n/(\Gamma_n+\Gamma_{a})^2\leq1/\Gamma_{a}$ for any value of the
(positive) escape width $\Gamma_n$.

\begin{figure}[b]
\includegraphics[width=0.475\textwidth]{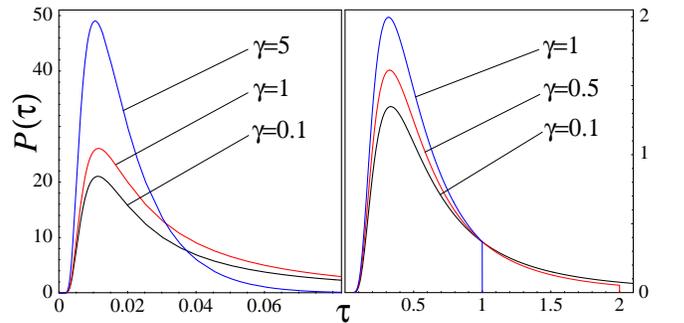} %
\caption{(color online). The distribution (\ref{Ptau1}) of the time delay in
single-channel scattering for different values of the absorption rate
$\gamma$ at weak ($T=0.1$, left) and perfect ($T=1$, right) coupling.}
\end{figure}
We analyze now the important case of single-channel scattering, $M=1$, in more
detail. The explicit expression to be obtained from (\ref{Ptau}) reads as
follows:
\begin{eqnarray}\label{Ptau1}
\mathcal{P}(\tau)&=&\frac{e^{-gs}}{\tau^{2}}\Bigl(I_0(\alpha\sqrt{g^2-1})
\bigl(\cosh\frac{\gamma}{2}-\frac{2}{\gamma}\sinh\frac{\gamma}{2}\bigr)
\nonumber\\ && %
+\frac{2}{\gamma}\sinh\frac{\gamma}{2}\bigl[gsI_0(\alpha\sqrt{g^2-1})
\nonumber\\ && %
- \alpha\sqrt{g^2-1}\,I_1(\alpha\sqrt{g^2-1})\bigr]\Bigr)\,.
\end{eqnarray}
We have explicitly checked the normalization of this distribution to unity
and verified relation (\ref{t-P}) for the first moment. This function should
be compared to the more simple expression
$\mathcal{P}_{\gamma=0}(\tau)=\tau^{-1}(\partial/\partial\tau)
e^{-g/\tau}I_0(\tau^{-1}\sqrt{g^2-1})$ \cite{Fyodorov1997} valid at zero
absorption. Figure 2 shows the behavior of $\mathcal{P}(\tau)$ in two
limiting cases of the weakly and perfectly open system. One sees in the first
case that the maximum of the distribution function at the small time
$\tau\sim(2g)^{-1}\approx T/4\ll1$ gets more pronounced and narrow as the
absorption rate $\gamma$ grows. At larger values of $\tau$ the distribution
is exponentially suppressed with
$\mathcal{P}(\gamma^{-1})\approx(\gamma^2/T)e^{-\gamma/T}$. The latter is
contrasted with the behavior in the case of perfect coupling, $g=T=1$, when
$\mathcal{P}_{T=1}(\tau)=\tau^{-2}e^{-1/\tau}[1+(1-\tau)(e^{\gamma}-1)/\tau\gamma]$
and $\mathcal{P}(\gamma^{-1})$ could be rather large. We relate these
distinctions to peculiarities in fluctuations of the resonance escape widths
in the two cases considered. The width distribution $\rho(\Gamma)$, which is
known exactly \cite{Fyodorov1997} for any $T$ and $M$, has the simple
exponential form $e^{-\Gamma t_H/T}$ when coupling is small, $T\ll1$, and the
power law behavior $\sim\Gamma^{-2}$ at $\Gamma\gtrsim t_H^{-1}$ when $T=1$.
The ratio of the widths $\Gamma\sim\Gamma_{a}$ determining
$\mathcal{P}(\tau\sim\gamma^{-1})$ is, therefore, exponentially small in the
first case and only a power in the second. This conclusion holds for any
finite $M$.

The sharp border at $\tau=\gamma^{-1}$ of the obtained distribution is the
direct consequence of equation (\ref{R}) with the absorption rate fixed to a
constant. Although, as shown above, the value $\mathcal{P}(\gamma^{-1})$ of
the jump may be exponentially small when coupling is weak, a generic
exponential suppression should be intuitively expected at large values of
delay times $\tau\gg\gamma^{-1}$. Indeed, for the time $\delta t$ a
wave-packet oscillating in the cavity with the frequency $\Delta/2\pi$ on
average experiences $(\Delta/2\pi)\delta t$ collisions with the walls,
yielding the probability $T_{\phi}(\Delta/2\pi)\delta t$ to be absorbed into
one of $M_{\phi}$ fictitious channels. The total reflection is then estimated
as $R\simeq(1-T_{\phi}(\Delta/2\pi)\delta t)^{M_{\phi}}$, giving
$e^{-\gamma\delta t/t_H}$ in the limit of fixed $\gamma=M_{\phi}T_{\phi}$ as
$M_{\phi}\to\infty$ and $T_{\phi}\to0$. It is instructive, therefore, to
define alternatively through the following relation  $R \equiv
e^{-\Gamma_{a}Q_R}$ the matrix $Q_R$, which we call the matrix of
\emph{reflection time-delays}. The positive definite matrix $Q_R$ is related
to $Q_{\gamma}$ as $Q_R =-\Gamma_{a}^{-1} \ln(1-\Gamma_{a}Q_{\gamma})$ that
leads to the following connection
\begin{equation}\label{PtauR}
\mathcal{P}_R(\tau_r) = e^{-\gamma\tau_r}
\mathcal{P}\bigl[\gamma^{-1}(1-e^{-\gamma\tau_r})\bigr]
\,,\quad \tau_r>0\,,
\end{equation}
between the corresponding distributions $\mathcal{P}_R(\tau_r)$ and
$\mathcal{P}(\tau)$ of proper delay times (eigenvalues of $Q_R$ and
$Q_{\gamma}$, respectively). Both $Q_R$ and $Q_{\gamma}$ reduce to the same
Wigner-Smith matrix (\ref{Q}) in the limit of vanishing absorption. The
difference between them becomes noticeable at finite $\gamma$. Still both
distributions coincide up to the time appreciably less than $\gamma^{-1}$.
They start to differ at larger times, when $\mathcal{P}(\tau)$ has the cutoff
whereas $\mathcal{P}_R(\tau\gg\gamma^{-1})\propto e^{-\gamma\tau}$ is
exponentially suppressed.

\begin{figure}[t]
\includegraphics[width=0.425\textwidth]{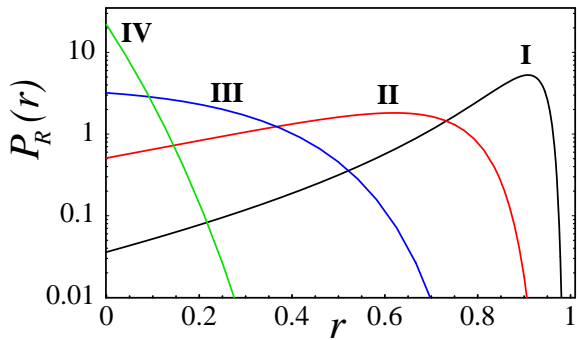} %
\caption{(color online). The reflection coefficient distribution in the
single-channel cavity at four experimental realizations \cite{reflection} of
the absorption rate and transmission coefficient (see the text for details).
The values $(2\gamma,T)$ correspond to I: (0.56, 0.12), II: (2.42, 0.75),
III: (8.4, 0.98), and IV: (48, 0.99).}
\end{figure}
Finally, we discuss the distribution $P_R(r)$ of the reflection coefficient
$r=|S_{\gamma}|^2=1-\gamma\tau$ in the single-channel cavity. This
distribution at arbitrary values of $\gamma$ and $T$ is explicitly given by
Eqs.~(\ref{P_R}) and (\ref{Ptau1}), reproducing exactly the recent result
\cite{yan} obtained by a different method. In the particular case of perfect
coupling it simplifies further to the expression
$P_{T=1}(r)=(1-r)^{-3}e^{-\gamma/(1-r)}[\gamma(e^{\gamma}-1)+(1+\gamma-e^{\gamma})(1-r)]$
found earlier \cite{Beenakker2001}. For the case of preserved TRS
($\beta=1$), the reflection coefficient distribution in a microwave cavity
has recently been measured \cite{reflection}. Our distribution $P_R(r)$ at
the values of absorption and transmission realized in this experiment under
compulsory (although not surprising in the RMT) rescaling $\gamma$ to
$\gamma\beta/2$ with $\beta=1$ is shown on Fig.~3. (This corresponds to
replacing our parameter $\gamma$ in (\ref{P_R}) and (\ref{Ptau1}) with
$T_w/2$ of Ref.~\cite{reflection}.) That should roughly take into account the
difference between the symmetry class of our analytical result ($\beta=2$)
and that of the experiment.  Such a replacement is expected to become more
efficient as absorption grows. The trend is clearly seen from the
distribution sharply peaked near $r\sim1$ at weak absorption ($\gamma\ll1$)
to the Rayleigh distribution $P_R(r)\simeq(\gamma\beta/2)
e^{-r\gamma\beta/2}$ \cite{Kogan2000}, see also \cite{Beenakker2001},
reproduced correctly at strong absorption ($\gamma\gg1$) and perfect coupling
($T=1$). Figure 3 is in good qualitative agreement with the experimental data
reported in Ref.~\cite{reflection} (see Figs.~4 and 6 there), which becomes
even quantitative as absorption gets stronger.  The rigorous analytical
treatment for the case of preserved TRS is still lacking, being under current
investigation.

In summary, we have calculated the general distribution of proper delay times
and reflection coefficients in an open chaotic system (e.g., billiard) with
broken TRS at arbitrary absorption. Finite absorption leads to strong
suppression of fluctuations at large values of delay times, making the
distribution narrower around the mean. The latter as well as the mean
reflection coefficient are found to be related to the ``norm leakage'' decay
function. The particular case of single-channel scattering is paid
appreciable attention, when discussion of available experimental data is also
given.

We are grateful to Y.V.~Fyodorov, G.~Hackenbroich, U.~Kuhl, V.V.~Sokolov,
H.-J.~St\"ockmann and C.~Viviescas for useful discussions. This work is
partly outcome from the ongoing initiative
Son\-der\-for\-schungs\-be\-reich/Transregio 6029. The partial financial
support by the SFB 237 and RFBR Grant No. 03-02-16151 (D.V.S.) is
acknowledged with thanks.


\begin{thebibliography}{35}
\expandafter\ifx\csname
natexlab\endcsname\relax\def\natexlab#1{#1}\fi
\expandafter\ifx\csname bibnamefont\endcsname\relax
  \def\bibnamefont#1{#1}\fi
\expandafter\ifx\csname bibfnamefont\endcsname\relax
  \def\bibfnamefont#1{#1}\fi
\expandafter\ifx\csname citenamefont\endcsname\relax
  \def\citenamefont#1{#1}\fi
\expandafter\ifx\csname url\endcsname\relax
  \def\url#1{\texttt{#1}}\fi
\expandafter\ifx\csname
urlprefix\endcsname\relax\def\urlprefix{URL }\fi
\providecommand{\bibinfo}[2]{#2}
\providecommand{\eprint}[2][]{\url{#2}}

\bibitem[{\citenamefont{Wigner}(1955)}]{Wigner1955}
\bibinfo{author}{\bibfnamefont{E.}~\bibnamefont{Wigner}},
  \bibinfo{journal}{Phys. Rev.} \textbf{\bibinfo{volume}{98}},
  \bibinfo{pages}{145} (\bibinfo{year}{1955}).

\bibitem[{\citenamefont{Smith}(1960)}]{Smith1960}
\bibinfo{author}{\bibfnamefont{F.~T.} \bibnamefont{Smith}},
  \bibinfo{journal}{Phys. Rev.} \textbf{\bibinfo{volume}{118}},
  \bibinfo{pages}{349} (\bibinfo{year}{1960}).

\bibitem{review}
\bibinfo{note}{For a general review of the concepts of the time delay and its
  applications see \bibinfo{author}{\bibfnamefont{C.~A.~A.}
  \bibnamefont{de~Carvalho}} \bibnamefont{and}
  \bibinfo{author}{\bibfnamefont{H.~M.} \bibnamefont{Nussenzveig}},
  \bibinfo{journal}{Phys. Rep.} \textbf{\bibinfo{volume}{364}},
  \bibinfo{pages}{83} (\bibinfo{year}{2002}) as well as
  Ref.~\cite{Fyodorov1997}}.

\bibitem[{\citenamefont{Sokolov and Zelevinsky}(1997)}]{Sokolov1997}
\bibinfo{author}{\bibfnamefont{V.~V.} \bibnamefont{Sokolov}} \bibnamefont{and}
  \bibinfo{author}{\bibfnamefont{V.}~\bibnamefont{Zelevinsky}},
  \bibinfo{journal}{Phys. Rev. C} \textbf{\bibinfo{volume}{56}},
  \bibinfo{pages}{311} (\bibinfo{year}{1997}).

\bibitem[{\citenamefont{Verbaarschot et~al.}(1985)\citenamefont{Verbaarschot,
  Weidenm{\"{u}}ller, and Zirnbauer}}]{Verbaarschot1985}
\bibinfo{author}{\bibfnamefont{J.~J.~M.} \bibnamefont{Verbaarschot}},
  \bibinfo{author}{\bibfnamefont{H.~A.} \bibnamefont{Weidenm{\"{u}}ller}},
  \bibnamefont{and} \bibinfo{author}{\bibfnamefont{M.~R.}
  \bibnamefont{Zirnbauer}}, \bibinfo{journal}{Phys. Rep.}
  \textbf{\bibinfo{volume}{129}}, \bibinfo{pages}{367} (\bibinfo{year}{1985}).

\bibitem[{\citenamefont{Beenakker}(1997)}]{Beenakker1997}
\bibinfo{author}{\bibfnamefont{C.~W.~J.} \bibnamefont{Beenakker}},
  \bibinfo{journal}{Rev. Mod. Phys.} \textbf{\bibinfo{volume}{69}},
  \bibinfo{pages}{731} (\bibinfo{year}{1997}).

\bibitem[{\citenamefont{Alhassid}(2000)}]{Alhassid2000}
\bibinfo{author}{\bibfnamefont{Y.}~\bibnamefont{Alhassid}},
  \bibinfo{journal}{Rev. Mod. Phys.} \textbf{\bibinfo{volume}{72}},
  \bibinfo{pages}{895} (\bibinfo{year}{2000}).

\bibitem[{\citenamefont{Lyuboshitz}(1977)}]{Lyuboshitz1977}
\bibinfo{author}{\bibfnamefont{V.~L.} \bibnamefont{Lyuboshitz}},
  \bibinfo{journal}{Phys. Lett. B} \textbf{\bibinfo{volume}{72}},
  \bibinfo{pages}{41} (\bibinfo{year}{1977});
\bibinfo{journal}{Yad. Fiz.} \textbf{\bibinfo{volume}{27}},
  \bibinfo{pages}{948} (\bibinfo{year}{1978}{\natexlab{b}})
  [{Sov. J. Nucl. Phys.} {\bf 27}, 502 (1978)].

\bibitem[{\citenamefont{Lehmann et~al.}(1995)\citenamefont{Lehmann, Savin,
  Sokolov, and Sommers}}]{Lehmann1995b}
\bibinfo{author}{\bibfnamefont{N.}~\bibnamefont{Lehmann}},
  \bibinfo{author}{\bibfnamefont{D.~V.} \bibnamefont{Savin}},
  \bibinfo{author}{\bibfnamefont{V.~V.} \bibnamefont{Sokolov}},
  \bibnamefont{and} \bibinfo{author}{\bibfnamefont{H.-J.}
  \bibnamefont{Sommers}}, \bibinfo{journal}{Physica D}
  \textbf{\bibinfo{volume}{86}}, \bibinfo{pages}{572} (\bibinfo{year}{1995}).

\bibitem[{\citenamefont{Fyodorov and Sommers}(1997)}]{Fyodorov1997}
\bibinfo{author}{\bibfnamefont{Y.~V.} \bibnamefont{Fyodorov}} \bibnamefont{and}
  \bibinfo{author}{\bibfnamefont{H.-J.} \bibnamefont{Sommers}},
  \bibinfo{journal}{J. Math. Phys.} \textbf{\bibinfo{volume}{38}},
  \bibinfo{pages}{1918} (\bibinfo{year}{1997});
\bibinfo{journal}{Phys. Rev. Lett.} \textbf{\bibinfo{volume}{76}},
  \bibinfo{pages}{4709} (\bibinfo{year}{1996}).

\bibitem[{\citenamefont{Gopar et~al.}(1996)\citenamefont{Gopar, Mello, and
  B{\"{u}}ttiker}}]{Gopar1996}
\bibinfo{author}{\bibfnamefont{V.~A.}~\bibnamefont{Gopar}},
  \bibinfo{author}{\bibfnamefont{P.~A.}~\bibnamefont{Mello}}, \bibnamefont{and}
  \bibinfo{author}{\bibfnamefont{M.}~\bibnamefont{B{\"{u}}ttiker}},
  \bibinfo{journal}{Phys. Rev. Lett.} \textbf{\bibinfo{volume}{77}},
  \bibinfo{pages}{3005} (\bibinfo{year}{1996}).

\bibitem[{\citenamefont{Fyodorov et~al.}(1997)\citenamefont{Fyodorov, Savin,
  and Sommers}}]{Fyodorov1997i}
\bibinfo{author}{\bibfnamefont{Y.~V.} \bibnamefont{Fyodorov}},
  \bibinfo{author}{\bibfnamefont{D.~V.} \bibnamefont{Savin}}, \bibnamefont{and}
  \bibinfo{author}{\bibfnamefont{H.-J.} \bibnamefont{Sommers}},
  \bibinfo{journal}{Phys. Rev. E} \textbf{\bibinfo{volume}{55}},
  \bibinfo{pages}{R4857} (\bibinfo{year}{1997});
\bibinfo{author}{\bibfnamefont{D.~V.} \bibnamefont{Savin}},
  \bibinfo{author}{\bibfnamefont{Y.~V.} \bibnamefont{Fyodorov}},
  \bibnamefont{and} \bibinfo{author}{\bibfnamefont{H.-J.}
  \bibnamefont{Sommers}}, \bibinfo{journal}{Phys. Rev. E}
  \textbf{\bibinfo{volume}{63}}, \bibinfo{pages}{035202(R)}
  (\bibinfo{year}{2001}).

\bibitem[{\citenamefont{Brouwer et~al.}(1997)\citenamefont{Brouwer, Frahm, and
  Beenakker}}]{Brouwer1997}
\bibinfo{author}{\bibfnamefont{P.~W.} \bibnamefont{Brouwer}},
  \bibinfo{author}{\bibfnamefont{K.~M.} \bibnamefont{Frahm}}, \bibnamefont{and}
  \bibinfo{author}{\bibfnamefont{C.~W.~J.} \bibnamefont{Beenakker}},
  \bibinfo{journal}{Phys. Rev. Lett.} \textbf{\bibinfo{volume}{78}},
  \bibinfo{pages}{4737} (\bibinfo{year}{1997});
\bibinfo{journal}{Waves Random Media} \textbf{\bibinfo{volume}{9}},
  \bibinfo{pages}{91} (\bibinfo{year}{1999}).

\bibitem[{\citenamefont{Sommers et~al.}(2001)\citenamefont{Sommers, Savin, and
  Sokolov}}]{Sommers2001}
\bibinfo{author}{\bibfnamefont{H.-J.} \bibnamefont{Sommers}},
  \bibinfo{author}{\bibfnamefont{D.~V.} \bibnamefont{Savin}}, \bibnamefont{and}
  \bibinfo{author}{\bibfnamefont{V.~V.} \bibnamefont{Sokolov}},
  \bibinfo{journal}{Phys. Rev. Lett.} \textbf{\bibinfo{volume}{87}},
  \bibinfo{pages}{094101} (\bibinfo{year}{2001}).

\bibitem[{\citenamefont{Doron et~al.}(1990)\citenamefont{Doron, Smilansky, and
  Frenkel}}]{Doron1990}
\bibinfo{author}{\bibfnamefont{E.}~\bibnamefont{Doron}},
  \bibinfo{author}{\bibfnamefont{U.}~\bibnamefont{Smilansky}},
  \bibnamefont{and} \bibinfo{author}{\bibfnamefont{A.}~\bibnamefont{Frenkel}},
  \bibinfo{journal}{Phys. Rev. Lett.} \textbf{\bibinfo{volume}{65}},
  \bibinfo{pages}{3072} (\bibinfo{year}{1990}).

\bibitem[{\citenamefont{Huibers et~al.}(1998)\citenamefont{Huibers, Patel,
  Marcus, Brouwer, Duru{\"o}z, and {Harris (Jr.)}}}]{Huibers1998b}
\bibinfo{author}{\bibfnamefont{A.~G.} \bibnamefont{Huibers}},
  \bibinfo{author}{\bibfnamefont{S.~R.} \bibnamefont{Patel}},
  \bibinfo{author}{\bibfnamefont{C.~M.} \bibnamefont{Marcus}},
  \bibinfo{author}{\bibfnamefont{P.~W.} \bibnamefont{Brouwer}},
  \bibinfo{author}{\bibfnamefont{C.~I.} \bibnamefont{Duru{\"o}z}},
  \bibnamefont{and}
  \bibinfo{author}{\bibfnamefont{J.~S.} \bibnamefont{{Harris, Jr.}}},
  \bibinfo{journal}{Phys. Rev. Lett.} \textbf{\bibinfo{volume}{81}},
  \bibinfo{pages}{1917} (\bibinfo{year}{1998}).

\bibitem[{\citenamefont{B{\"{u}}ttiker}(1986)}]{Buettiker1986}
\bibinfo{author}{\bibfnamefont{M.}~\bibnamefont{B{\"{u}}ttiker}},
  \bibinfo{journal}{Phys. Rev. B} \textbf{\bibinfo{volume}{33}},
  \bibinfo{pages}{3020} (\bibinfo{year}{1986}).

\bibitem[{\citenamefont{Alves and Lewenkopf}(2002)}]{Alves2002}
\bibinfo{author}{\bibfnamefont{E.~R.~P.} \bibnamefont{Alves}} \bibnamefont{and}
  \bibinfo{author}{\bibfnamefont{C.~H.} \bibnamefont{Lewenkopf}},
  \bibinfo{journal}{Phys. Rev. Lett.} \textbf{\bibinfo{volume}{88}},
  \bibinfo{pages}{256805} (\bibinfo{year}{2002}).

\bibitem[{\citenamefont{Baranger and Mello}(1995)}]{Baranger1995}
\bibinfo{author}{\bibfnamefont{H.~U.} \bibnamefont{Baranger}} \bibnamefont{and}
  \bibinfo{author}{\bibfnamefont{P.~A.} \bibnamefont{Mello}},
  \bibinfo{journal}{Phys. Rev. B} \textbf{\bibinfo{volume}{51}},
  \bibinfo{pages}{4703} (\bibinfo{year}{1995}).

\bibitem[{\citenamefont{Brouwer and Beenakker}(1997)}]{Brouwer1997ii}
\bibinfo{author}{\bibfnamefont{P.~W.} \bibnamefont{Brouwer}} \bibnamefont{and}
  \bibinfo{author}{\bibfnamefont{C.~W.~J.} \bibnamefont{Beenakker}},
  \bibinfo{journal}{Phys. Rev. B} \textbf{\bibinfo{volume}{55}},
  \bibinfo{pages}{4695} (\bibinfo{year}{1997}) [Erratum: Phys.
  Rev. B \textbf{66}, 209901(E) (2002)].

\bibitem[{\citenamefont{Ramakrishna and Kumar}(2000)}]{Ramakrishna2000}
\bibinfo{author}{\bibfnamefont{S.~A.} \bibnamefont{Ramakrishna}}
  \bibnamefont{and} \bibinfo{author}{\bibfnamefont{N.}~\bibnamefont{Kumar}},
  \bibinfo{journal}{Phys. Rev. B} \textbf{\bibinfo{volume}{61}},
  \bibinfo{pages}{3163} (\bibinfo{year}{2000}).

\bibitem[{\citenamefont{Beenakker and Brouwer}(2001)}]{Beenakker2001}
\bibinfo{author}{\bibfnamefont{C.~W.~J.} \bibnamefont{Beenakker}}
  \bibnamefont{and} \bibinfo{author}{\bibfnamefont{P.~W.}
  \bibnamefont{Brouwer}}, \bibinfo{journal}{Physica E}
  \textbf{\bibinfo{volume}{9}}, \bibinfo{pages}{463} (\bibinfo{year}{2001}).

\bibitem[{\citenamefont{Sch{\"{a}}fer et~al.}(2003)\citenamefont{Sch{\"{a}}fer,
  Gorin, Seligman, and St{\"o}ckmann}}]{Schaefer2003}
\bibinfo{author}{\bibfnamefont{R.}~\bibnamefont{Sch{\"{a}}fer}},
  \bibinfo{author}{\bibfnamefont{T.}~\bibnamefont{Gorin}},
  \bibinfo{author}{\bibfnamefont{T.~H.} \bibnamefont{Seligman}},
  \bibnamefont{and} \bibinfo{author}{\bibfnamefont{H.-J.}
  \bibnamefont{St{\"o}ckmann}},
  \bibinfo{journal}{J. Phys. A: Math. Gen.} \textbf{\bibinfo{volume}{36}},
  \bibinfo{pages}{3289} (\bibinfo{year}{2003}).

\bibitem[{\citenamefont{Bohr and Mottelson}(1969)}]{Bohr1969}
\bibinfo{author}{\bibfnamefont{A.}~\bibnamefont{Bohr}} \bibnamefont{and}
  \bibinfo{author}{\bibfnamefont{B.~R.} \bibnamefont{Mottelson}},
  \emph{\bibinfo{title}{Nuclear Structure}} (\bibinfo{publisher}{Benjamin},
  \bibinfo{address}{New York}, \bibinfo{year}{1969}), Vol.~1.

\bibitem{models}
\bibinfo{note}{It is worth noting that the case of absorption, when
  dissolution in walls occurs, should be contrasted with that of dephasing,
  when the particle number in the cavity is conserved; see \cite{Brouwer1997ii}}.

\bibitem[{\citenamefont{Savin and Sokolov}(1997)}]{Savin1997}
\bibinfo{author}{\bibfnamefont{D.~V.} \bibnamefont{Savin}} \bibnamefont{and}
  \bibinfo{author}{\bibfnamefont{V.~V.} \bibnamefont{Sokolov}},
  \bibinfo{journal}{Phys. Rev. E} \textbf{\bibinfo{volume}{56}},
  \bibinfo{pages}{R4911} (\bibinfo{year}{1997}).

\bibitem{shift}
\bibinfo{note}{Only in this case the smooth energy dependence of the coupling
  amplitudes can be neglected, so that any scattering quantity depends
  explicitly on $E$ by means of $(E-\mathcal{H})^{-1}$ only}.

\bibitem[{\citenamefont{Kogan et~al.}(2000)\citenamefont{Kogan, Mello, and
  Liqun}}]{Kogan2000}
\bibinfo{author}{\bibfnamefont{E.}~\bibnamefont{Kogan}},
  \bibinfo{author}{\bibfnamefont{P.~A.} \bibnamefont{Mello}}, \bibnamefont{and}
  \bibinfo{author}{\bibfnamefont{H.}~\bibnamefont{Liqun}},
  \bibinfo{journal}{Phys. Rev. E} \textbf{\bibinfo{volume}{61}},
  \bibinfo{pages}{R17} (\bibinfo{year}{2000}).

\bibitem[{\citenamefont{Efetov}(1996)}]{Efetov1996}
\bibinfo{author}{\bibfnamefont{K.~B.} \bibnamefont{Efetov}},
  \emph{\bibinfo{title}{Supersymmetry in Disorder and Chaos}}
  (\bibinfo{publisher}{Cambridge University Press},
  \bibinfo{address}{Cambridge, UK}, \bibinfo{year}{1996}).

\bibitem{misprint}
\bibinfo{note}{
  This is the usual procedure in the crossover regime of gradually
  broken TRS; see \cite{Fyodorov1997i,Efetov1996}.
  Note also a misprint in \cite{Savin1997}:
  $f(\lambda_i)=(\lambda_1^2-\lambda^2)/4$ is correct}.

\bibitem[{\citenamefont{Fyodorov and Alhassid}(1998)}]{Fyodorov1998i}
\bibinfo{author}{\bibfnamefont{Y.~V.} \bibnamefont{Fyodorov}} \bibnamefont{and}
  \bibinfo{author}{\bibfnamefont{Y.}~\bibnamefont{Alhassid}},
  \bibinfo{journal}{Phys. Rev. A} \textbf{\bibinfo{volume}{58}},
  \bibinfo{pages}{R3375} (\bibinfo{year}{1998}).

\bibitem{susy}
\bibinfo{note}{The generating function (4) of \cite{Sommers2001} can be
  again written in terms of the $2N\!\times\!2N$ determinants
  $\mathrm{Det}(A_{\pm})$,
  where now
  $A_{\pm}=\frac{1}{2}VV^{\dagger}-(z_{\pm}^{-1}-\frac{1}{2}\Gamma_{a})
  -z_{\pm}^{-1}\sqrt{1-z_{\pm}\Gamma_{a}}\,\sigma_1-i(E\!-\!H)\sigma_3$}.

\bibitem{table}
\bibinfo{note}{$\int_{1}^{\infty}\!\!dt\,e^{-st}I_0[\pm\alpha\sqrt{t^2-1}]
 =e^{\sqrt{s^2-\alpha^2}}/\sqrt{s^2-\alpha^2}$ for $s>\alpha\geq0$.
 For the calculation of the imaginary part we use:
 $\int_{-1}^{1}\!\!dt\,e^{-\alpha gt}\cos(\alpha\sqrt{1-t^2})/\sqrt{1-t^2}=
 \pi I_0[\alpha\sqrt{g^2-1}]$}.

\bibitem{tail}
\bibinfo{note}{Similar suppression was previously observed in
  Ref.~\cite{Ramakrishna2000} for the time-delay distribution in single-channel
  reflection from one-dimensional amplifying random media}.

\bibitem{yan}
\bibinfo{author}{\bibfnamefont{Y.~V.} \bibnamefont{Fyodorov}},
   cond-mat/0304671.

\bibitem{reflection}
\bibinfo{author}{\bibfnamefont{R.~A.} \bibnamefont{M\'endez-S\'anchez}
  \textit{et al.} } cond-mat/0305090.

\end{thebibliography}

\end{document}